\documentclass[a4paper]{article}

\usepackage{INTERSPEECH_v2}

\usepackage{graphicx}
\usepackage{color}
\usepackage{amssymb,amsmath,bm}
\usepackage{textcomp}

\def\vec#1{\ensuremath{\bm{{#1}}}}

\ninept

\title{Generative adversarial network-based glottal waveform model for statistical parametric speech synthesis}

\makeatletter
\def\name#1{\gdef\@name{#1\\}}
\makeatother \name{{\em Bajibabu Bollepalli, Lauri Juvela, Paavo Alku}}

\address{Department of Signal Processing and Acoustics, Aalto University, Finland
}
\email{firstname.lastname@aalto.fi}
\begin{document}

  \maketitle
  \begin{abstract}
  Recent studies have shown that text-to-speech synthesis quality can be improved by using glottal vocoding.
 This refers to vocoders that parameterize speech into two parts, the glottal excitation and vocal tract, that occur in the human speech production apparatus. Current glottal vocoders generate the glottal excitation waveform by using deep neural networks (DNNs). However, the squared error-based training of the present glottal excitation models is limited to generating conditional average waveforms, which fails to capture the stochastic variation of the waveforms. As a result, shaped noise is added as post-processing. 
 In this study, we propose a new method for predicting glottal waveforms by generative adversarial networks (GANs).
 GANs are generative models that aim to embed the data distribution in a latent space, enabling generation of new instances very similar to the original by randomly sampling the latent distribution.
The glottal pulses generated by GANs show a stochastic component similar to natural glottal pulses. In our experiments, we compare synthetic speech generated using glottal waveforms produced by both DNNs and GANs.
  The results show that the newly proposed GANs achieve synthesis quality comparable to that of widely-used DNNs, without using an additive noise component.
  \end{abstract}

  \section{Introduction} \label{sec:intro}
 Statistical parametric speech synthesis (SPSS) and concatenative synthesis are the two predominant paradigms in text-to-speech technology. SPSS systems have several advantages over concatenative synthesis, such as their flexibility to transform the synthesis to different voice characteristics,
 speaking styles and emotions, as well as their small memory footprint and robustness to unseen text prompts \cite{Tokuda2013}. The main drawback of SPSS is that the quality  of synthetic speech is worse than that of concatenative synthesis. There are three major factors behind this: quality of vocoders, acoustic modeling accuracy, and over-smoothing \cite{Zen2009-SPSS}.
  
  Recent use of neural network-based acoustic models \cite{Zen2013-DNN}, especially sequence models, such as long short-term memory (LSTM) networks \cite{fan2014tts,zen2015unidirectional}, have addressed primarily the acoustic modelling accuracy and to some extent also the over-smoothing problem \cite{wu2016improving,wang2017-ar-mdn}. Although the progress in acoustic modelling has improved the synthesis, the quality achieved by the best SPSS systems is still limited by the copy-synthesis quality of the vocoder. Thus in this study we focus on improving the quality of vocoders.
  
  In SPSS systems, vocoders are used for speech parametrization and waveform generation. Vocoders used in SPSS can be grouped into three main categories: mixed/impulse excited vocoders (e.g. STRAIGHT \cite{Kawahara1999,Kawahara2001}), glottal vocoders (e.g. GlottHMM \cite{Raitio2011glotthmm} and GlottDNN \cite{airaksinen2016glottdnn}), and sinusoidal vocoders (e.g. quasi harmonic model \cite{stylianou2001applying,erro2014vocoder}). The first two categories are based on the source-filter model of speech production and they differ mainly on the interpretation of voiced excitation signal. In glottal vocoding, the excitation is assumed to correspond to the time-derivative of the true airflow generated at the vocal folds (consisting of the combined effects of the glottal volume velocity and lip radiation \cite{Alku1992}),
  and the filter corresponds to a transfer function that is created by the
  %physiological organs
  physiology
  of the human vocal tract.
  Recent studies have shown that glottal vocoding can improve the synthesis quality \cite{Raitio2011glotthmm,juvela2016a-high-pitched-excitation,Airaksinen2017}.
  
The first glottal vocoder \cite{Raitio2011glotthmm} used a single glottal pulse to create the voiced excitation waveform---the pulse was modified according to the estimated acoustic parameters to build the entire excitation signal. This straightforward use of a single glottal pulse was replaced in later studies \cite{Raitio2014c-deep-neural-network-based-trainable,juvela2016a-high-pitched-excitation} with deep neural networks (DNNs) to predict the glottal pulse waveforms from acoustic features, where the actual estimated glottal pulses
were set as optimization targets.
%are employed in the network optimization.
However, the mean squared error-based training of the DNN-based glottal excitation models is only able to generate conditional average waveforms, which fails to capture the stochastic variation of the waveforms. In order to tackle this drawback, the excitation was post-processed by adding shaped noise to the waveform. In this study, we propose an alternative method for predicting glottal excitation waveforms for SPSS, by using a new training strategy with generative adversarial networks (GANs) \cite{goodfellow2014generative}.  
  
 As the research in deep learning progresses, new advanced neural networks capable of generating raw signal waveforms directly from linguistic features are proposed in text-to-speech (e.g WaveNet \cite{oord2016-wavenet}, Deep Voice \cite{Arik2017-deepvoice}, and \cite{hega2017gen}).
Although these systems produce high-quality synthetic speech they are not yet applicable in real-time speech synthesis due to heavy computational requirements. In contrast, the widely-used source-filter model is an applicable means to express speech in parametric forms as shown by its widespread applications \cite{drugman2014glottal,Alku2011}. In addition, the excitation of the source-filter model, the glottal flow, is an elementary time-domain signal (particularly when compared to the speech pressure signal) because it is produced at the level of glottis in the human larynx in the absence of vocal tract resonances. Hence, the glottal excitation is an attractive domain for generative waveform modeling.

%GANs have been previously used only in two previous TTS studies.
Recently, GANs have started to emerge in TTS applications.
In \cite{Kaneko2017-gan-postfilter}, a GAN was employed as a post-filter to address the over-smoothing problem in predicted acoustic parameters in SPSS. Results of \cite{Kaneko2017-gan-postfilter} showed that GANs are capable of producing detailed speech spectra, including also the modulation spectrum, resulting in increased synthesis quality. 
In \cite{Yuki2017}, adversarial type of training was used to take into account an anti-spoofing verification as an additional constraint in the acoustic model training. The current study is the first investigation to use GANs to model the glottal waveform as an excitation waveform in SPSS.
  
\section{Generative adversarial networks} \label{sec:gans}
   \begin{figure}[t]
        \centering
        \includegraphics[width=0.9\linewidth, height=4cm]{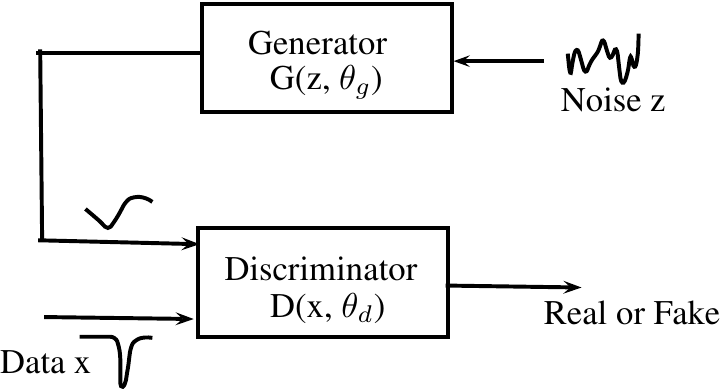}
        \caption{{\it General block diagram of generative adversarial networks (GANs).}}
        \label{fig:gan_block}
      \end{figure}

Generative adversarial networks (GANs) are generative models that have shown a huge success in unsupervised learning \cite{radford2015unsupervised}. In GANs a new type of training procedure is employed by an adversarial process where two models, generator $G$ and discriminator $D$, compete with each other.
%The first model is a generator $G$ and the second model is a discriminator $D$.
Figure~\ref{fig:gan_block} illustrates the block diagram of a GAN.  During training, $G$ starts from sampling input variables $\vec{z}$ from a uniform or Gaussian distribution $p_{\vec{z}}(\vec{z})$, then maps the input latent variables $\vec{z}$ to data space $G(\vec{z}; \theta_g)$ through a differentiable network. $D$ is a classifier $D(\vec{x}; \theta_d)$ that aims to discriminate whether a sample is a real one from the training data or a fake generated by $G$.  In this framework, $D$ and $G$ play a two-player minmax game with the following binary cross entropy:

  \begin{equation}
    \label{eq_gan}
    \begin{split}
        \min_G \max_D V_{\text{\tiny GAN}}(D, G) = \mathbb{E}_{\bm{x} \sim p_{\text{data}}(\bm{x})}[\log D(\bm{x})]+\\ \mathbb{E}_{\bm{z} \sim p_{\bm{z}}(\bm{z})}[\log (1 - D(G(\bm{z})))].
    \end{split}
 \end{equation}

In training, updates are alternated between $G$ and $D$, but the error gradient always propagates through the classifier $D$. 
The main theoretical advantage of this framework is that the parameters $\theta_g$ and $\theta_d$ can be learned through back propagation without making any assumptions on the data distribution \cite{goodfellow2014generative}.
  
\section{GAN-based glottal waveform model} \label{sec:gan_gw}
The regular or vanilla GAN \cite{goodfellow2014generative} framework is modified in the following manner to model glottal waveforms.
 
 \subsection{Conditional generative adversarial networks (CGAN)} \label{subsec:cgan}
 Generator $G$ in regular GANs has no control on modes of data it generates. In \cite{mirza2014conditional} it was shown that by conditioning the model on additional information it is possible to direct the data generation process. Since the goal of the current study is to generate glottal pulses based on acoustic parameters, we conditioned both the generator and discriminator by the acoustic parameters $\vec{y}$. The objective function in Eq.~\ref{eq_gan} can be rewritten with conditional variable $\vec{y}$ as:
 
 \begin{equation}
    \label{eq:minimaxgame-definition-conditioned}
    \begin{split}
    \min_G \max_D V(D, G) = \mathbb{E}_{\bm{x} \sim p_{\text{data}}(\bm{x})}[\log D(\bm{x} | \bm{y})] + \\ \mathbb{E}_{\bm{z} \sim p_z(\bm{z})}[\log (1 - D(G(\bm{z} | \bm{y})))].
    \end{split}
\end{equation}
 
 \subsection{Convolutional architecture} \label{subsec:dcgan}
 In regular GANs, both discriminator and generator employ a simple feed-forward neural network for learning. However, numerous studies have shown (e.g.~\cite{radford2015unsupervised,fu2017raw} that convolutional architectures yield better generated outputs than simple feed-forward networks. %in image generation tasks.
Our generator network consists of only convolutional layers and hence the local temporal characteristics in glottal waveforms can be effectively preserved with a relatively small number of weights.
  
 \subsection{Least Squares Generative Adversarial Networks} \label{subsec:lsgan}
 In the regular GAN, the discriminator is a classifier and uses binary cross-entropy as a loss function. In \cite{Mao2017lsgan} it was shown that this kind of loss function can lead to problems due to vanishing gradients when updating the parameters of the generator. Thus the loss function of discriminator in Eq.~\ref{eq:minimaxgame-definition-conditioned} is modified to the least square function:
   \begin{equation}
  \label{eq:lsgan_01}
  \begin{split}
  \min_D V_{\text{\tiny LSGAN}}(D) = \frac{1}{2}\mathbb{E}_{\bm{x} \sim p_{\text{data}}(\bm{x})}\bigl[(D(\bm{x | \bm{y}})-1)^2\bigr] \\
  + \frac{1}{2}\mathbb{E}_{\bm{z} \sim p_{\bm{z}}(\bm{z})}\bigl[(D(G(\bm{z} | \bm{y})))^2\bigr]
  %\min_G V_{\text{\tiny LSGAN}}(G) = &\frac{1}{2}\mathbb{E}_{\bm{z} \sim p_{\bm{z}}(\bm{z})}\bigl[(D(G(\bm{z}))-1)^2\bigr],
  \end{split}
  \end{equation}

\section{Experiments} \label{sec:experiments}
   \begin{figure}[t]
        \centering
        \includegraphics[width=0.9\linewidth, height=7.5cm]{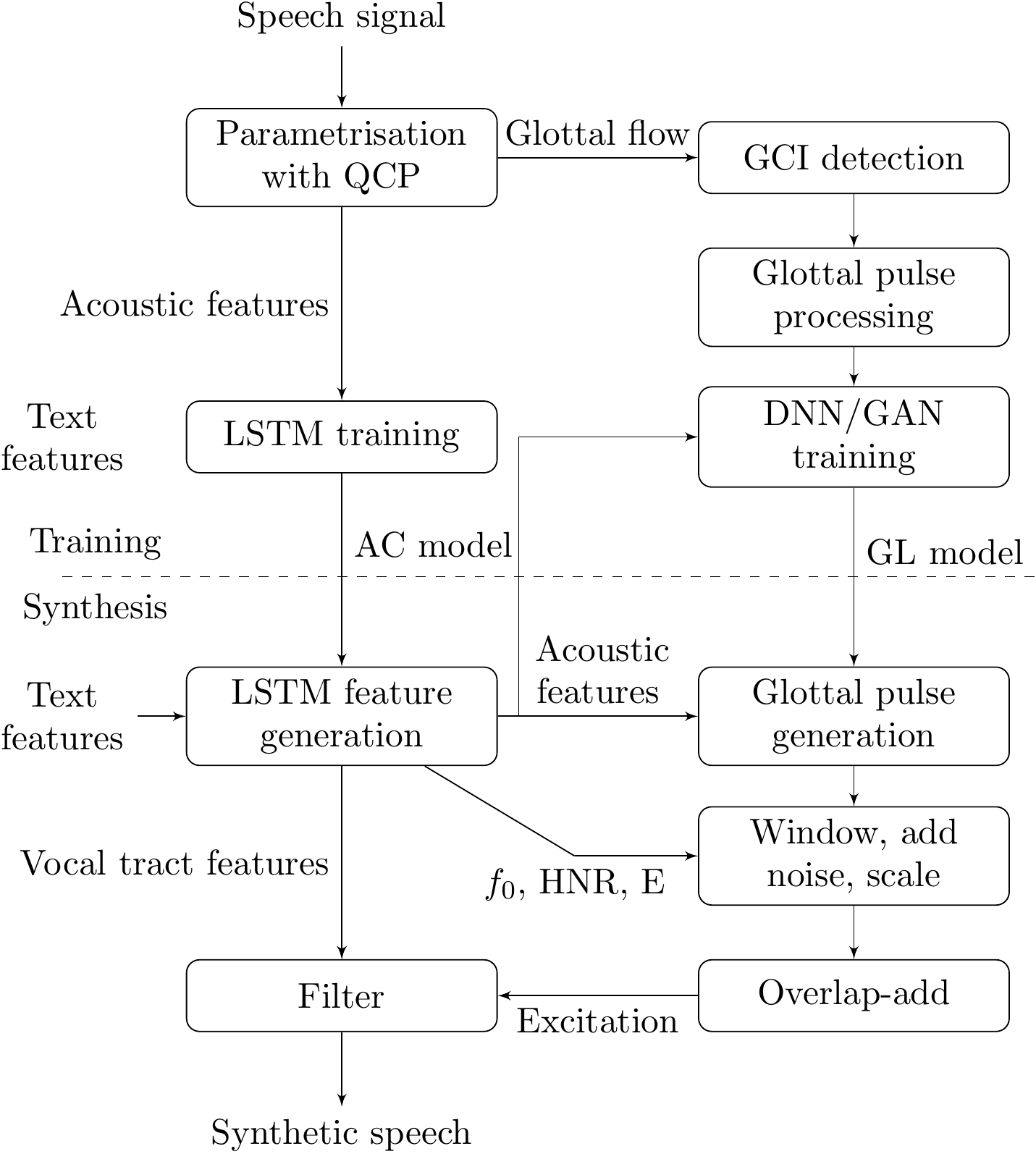}
        \caption{{\it Block diagram of the LSTM-based speech synthesis system using the GlottDNN vocoder.}}
        \label{fig:synthesis_blockdiagram}
      \end{figure}
      
\subsection{Speech material} \label{subsec:speechmaterial}
 We employed data of one female speaker recorded by a professional British English voice talent, labeled as ``Jenny''. The speech data consisted of 4314 utterances summing to 7 hours and 51 minutes. A total of 100 utterances were randomly selected for validation and testing, and the rest were used for training the systems. The sampling frequency of the corpus was 16 kHz.
 
 \subsection{Feature extraction} \label{subsec:featext}
 \subsubsection{Linguistic features}
 Figure~\ref{fig:synthesis_blockdiagram} illustrates the block diagram of our LSTM-based speech synthesis system used in this study. The text files of the utterances were provided along with corpus. The full contextual labels were obtained by the Flite \cite{black2001flite} speech synthesis front end and the Combilex \cite{richmond2009combilex} lexicon. To align the labels and acoustic features at phoneme level, the HMM-based force alignment was used. The full-context labels were represented into binary and numerical features by the question file used in the HMM-based speech synthesis system. These features convey information about the phoneme identity, syllable location, part-of-speech, number of words in an utterance, and number of phrases in an utterance.
In total, the input feature vector was 396 in dimension (per time-frame) including the extra numerical values which provide information about the frame position in a given phoneme.
 
 \subsubsection{Acoustic features}
 Both the vocal tract and voice source parameters, shown in Table~\ref{tab:params}, were extracted using the GlottDNN vocoder \cite{airaksinen2016glottdnn}. The acoustic parameters were extracted at a 5-ms frame rate. The $\log F0$  was linearly interpolated to fill unvoiced regions and an extra binary V/UV feature was added to code the voiced/unvoiced 
information. The output parameters included both static and dynamic (with delta, and delta-delta) features. Thus, in total, the output feature was 142 dimensional. The input features were normalized to the range of [0.1, 0.99] by using the min-max method, while output features were normalized using the mean-variance normalization method. The development and evaluation set were normalized by the values derived from the training data. At synthesis time,  the maximum likelihood parameter generation (MLPG) \cite{zen2004introduction} algorithm was applied on predicted acoustic parameters using the global variances to generate smooth parameter trajectories.
 
 \subsection{Acoustic (AC) model}
 The acoustic model network consisted of four hidden layers which were followed by a linear layer at the output. The four hidden layers consisted of two feed-forward layers at bottom and two bidirectional LSTM layers on top. The bottom feed-forward layers were intended to act as feature extraction layers, with 512 hidden units using logistic activation function in each layer. The top two layers had 256 bidirectional LSTM blocks in each layer. The stochastic gradient descent algorithm was used to learn the parameters and early stopping criterion was adopted to reduce overfitting.
 \begin{table}[tb]
      \caption{Acoustic features used in training the LSTM-based AC model, the DNN and the GAN-based GL model.}
     \label{tab:params}
     \centering
        \begin{tabular}{|c|c|c|}
            \hline
             Feature                 & Type/Unit &  Dimension  \\ \hline
             Vocal tract spectrum    & LSF       &  30        \\
             Energy                  & dB        &  1          \\
             Fundamental frequency   & $\log F_0$  &  1          \\
             Harmonic-to-noise ratio & dB/ERB    &  5          \\
             Voice source spectrum   & LSF       &  10         \\ \hline
             Total                   & -         & 47          \\ \hline
        \end{tabular}
 \end{table}

\subsection{Glottal excitation (GL) model}
\subsubsection{DNN-based glottal excitation model}
The GL model using DNNs was developed as described in \cite{juvela2016a-high-pitched-excitation}. The input features to the network were the same acoustic features as described in Table~\ref{tab:params}  (i.e.~47 in dimension) and the outputs were two pitch-period windowed glottal flow derivative waveforms centered and zero-padded to 400 time-domain samples. The acoustic parameters predicted by the AC model were employed to train the GL model instead of the original acoustic features, in contrast to \cite{juvela2016a-high-pitched-excitation}. The main motivation for this change is to reduce the mismatch in the acoustic feature inputs between training and testing time -- we provide a detailed analysis on this issue in \cite{juvela2017a-reduce-mismatch}. The DNN architecture consisted of three hidden layers each with 512 units. The logistic and linear activations were used for hidden and output layers, respectively.

\subsubsection{GAN-based glottal excitation model}
\begin{figure}[htb]
\begin{minipage}[b]{.48\linewidth}
  \centering
  \centerline{\includegraphics[width=3.9cm, height=4.0cm]{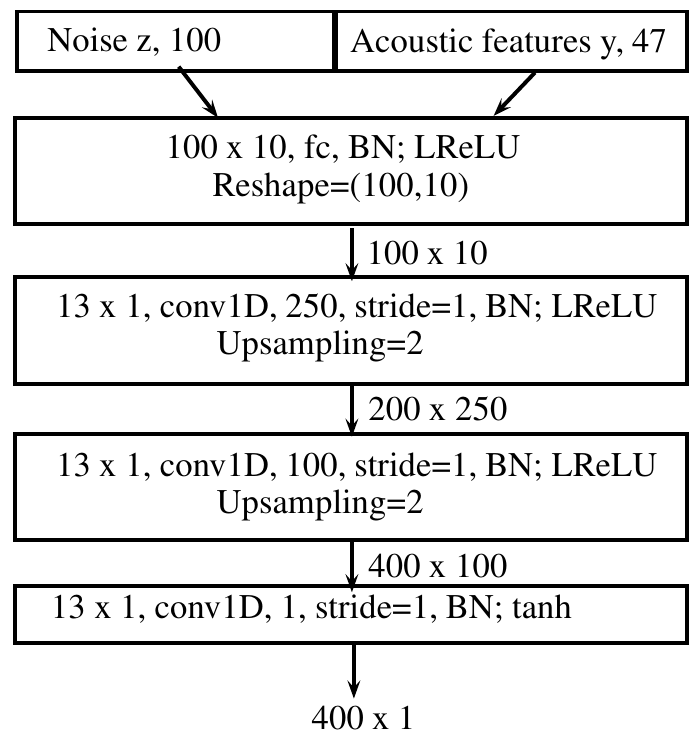}}
%  \vspace{2.0cm}
  \centerline{(a) Generator}\medskip
\end{minipage}
%\hspace{-0.5cm}
\begin{minipage}[b]{.48\linewidth}
  \centering
  \centerline{\includegraphics[width=3.9cm,height=4.0cm]{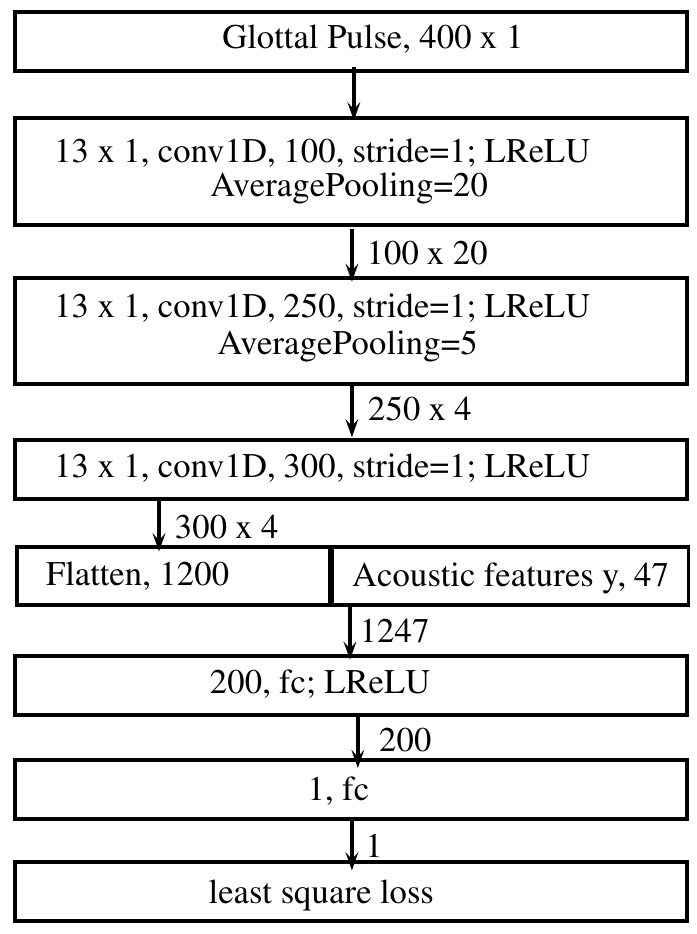}}
%  \vspace{1.5cm}
  \centerline{(b) Discriminator}\medskip
\end{minipage}
%\vspace{-0.5cm}
\caption{ Architectures of models CGAN+CNN and CGAN+CNN+LS. BN: batch normalization, fc: fully connected layer, conv1D: 1D convolution.}
\label{fig:cgan_arc}
\end{figure}

Four types of GL models were developed using GANs. 
The first model was a vanilla GAN, denoted as ``GAN'', where the DNNs were employed for both the generator and discriminator. The generator consisted of three hidden layers followed by an output layer. The discriminator consisted
four hidden layers followed by an output layer. Each hidden layer had 1024 units and the activation function was a leaky rectified linear unit (LReLU). The tangent hyperbolic and sigmoid activation functions were employed in the output layer of the generator and discriminator, respectively. The batch-normalization was employed on the generator network \cite{ioffe2015batch}. The second model was a conditional GAN, denoted as ``CGAN'', same as the vanilla GAN except that it was conditioned by acoustic features in both the generator and discriminator. The third model, denoted as ``CGAN+CNN'', was a conditional GAN with deep convolutional neural networks in both the generator and discriminator. The fourth model, denoted as ``CGAN+CNN+LS'', was similar to the third model except that the least square loss was used in the discriminator. Architectures of the third and fourth models are illustrated in Figure \ref{fig:cgan_arc}. The noise vector $\vec{z}$ had a dimension of 100 and was sampled from Gaussian distribution $\mathcal{N}(0,0.5)$.

%\vspace{-0.5cm}

      \begin{figure}[t]
        \centering
        \includegraphics[width=0.9\linewidth, height=6.5cm]{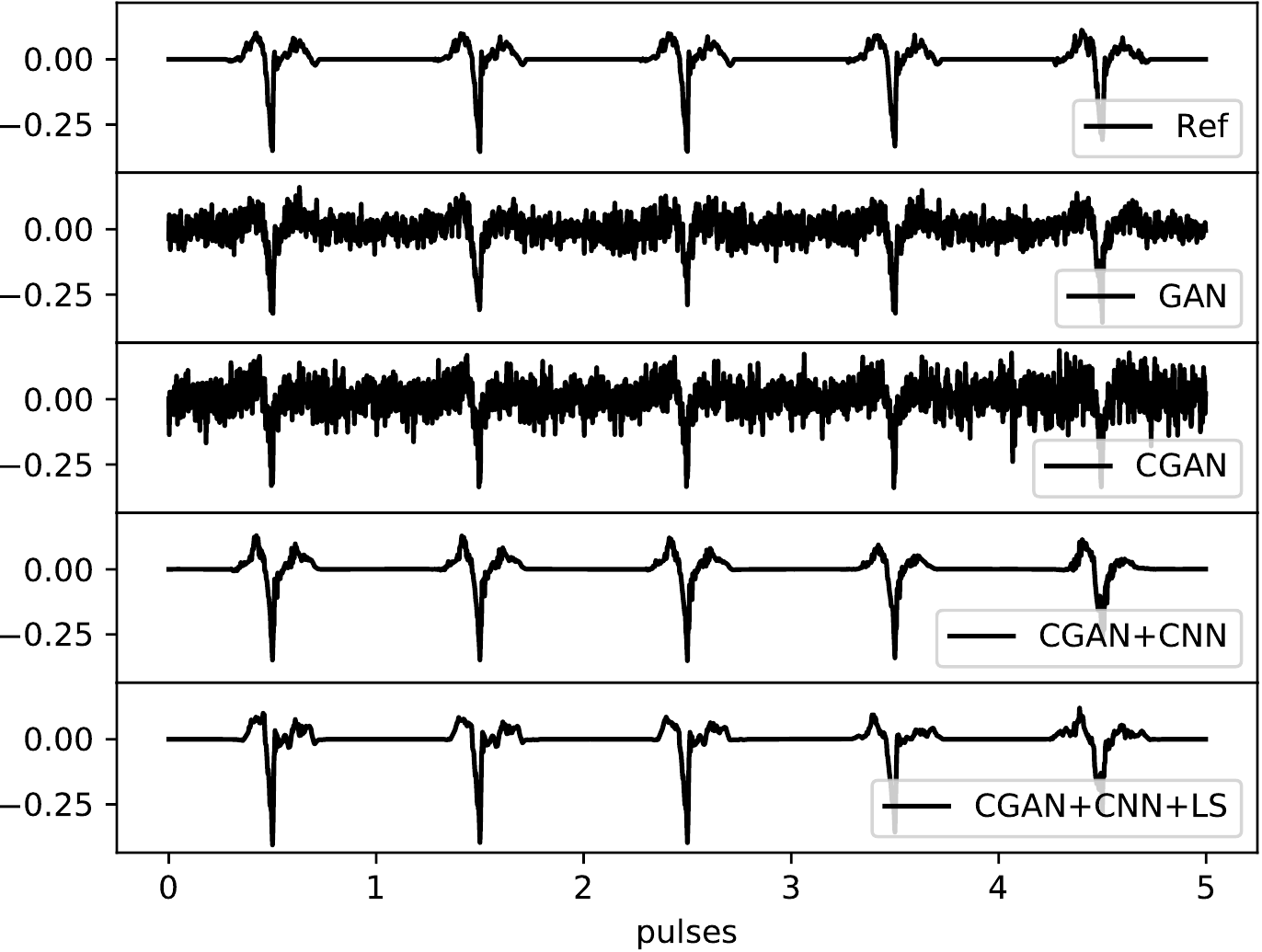}
        \caption{{\it Glottal pulses generated by different generative adversarial networks (GANs). Ref: natural reference, GAN: vanilla GANs, CGAN: conditional GAN, CGAN+CNN: conditional GAN with deep convolutional neural networks, and CGAN+CNN+LS: same as CGAN+CNN but least square loss is used by the discriminator.}}
        \label{fig:gans_pulses}
      \end{figure}

\subsection{Objective evaluation}

% lauri: Both MSE and Pearson correlations are calculated point-wise, which is a bit problematic -- even if GAN captures the correct noise distribution, the generated noise realizations will not match the reference pulses point-wise
% We should rather use some spectrum-based measures, like MFCC distortion or power-spectrum KL-divergence

%Since the aim of the study was to generate a glottal pulse resemble to natural glottal pulse using GANs, 
The main drawback of GANs is the lack of an explicit objective score to measure the performance of the generator \cite{goodfellow2014generative}. Therefore, visual inspection is typically adopted \cite{radford2015unsupervised}.  In the current study, simple objective scores, the mean square error (MSE) and Pearson correlation coefficient (PCC) computed between the actual and generated glottal pulses, were used. The obtained objective scores, computed as an average over 100 utterances (with 43746 glottal pulses), are presented in Table \ref{tab:objscores}.

%We conducted objective evaluations across the four proposed models and also the baseline DNN GL model. 

The MSE value of the baseline DNN system was lower than that of the other systems, but this was expected since the DNN system was trained to minimize the MSE cost function. Among the proposed models, deep convolutional neural network (CNN) -based GAN models outperformed the DNN-based GAN models in both MSE and PCC. Figure~\ref{fig:gans_pulses} shows a few example pulses generated by the proposed methods. It can be seen that the glottal pulses generated by the CNN-based GAN models are visually much closer to the reference pulses than the corresponding pulses generated by the DNN-based GAN. 
 %For Pulses in the
 % We have observed that LSGANs are able to generate higher quality pulses than vanilla GANs. And also LSGANs performs more stable during training. Thus the loss function is modified to 
     \begin{figure}[t]
        \centering
        \includegraphics[width=0.9\linewidth, height=6cm]{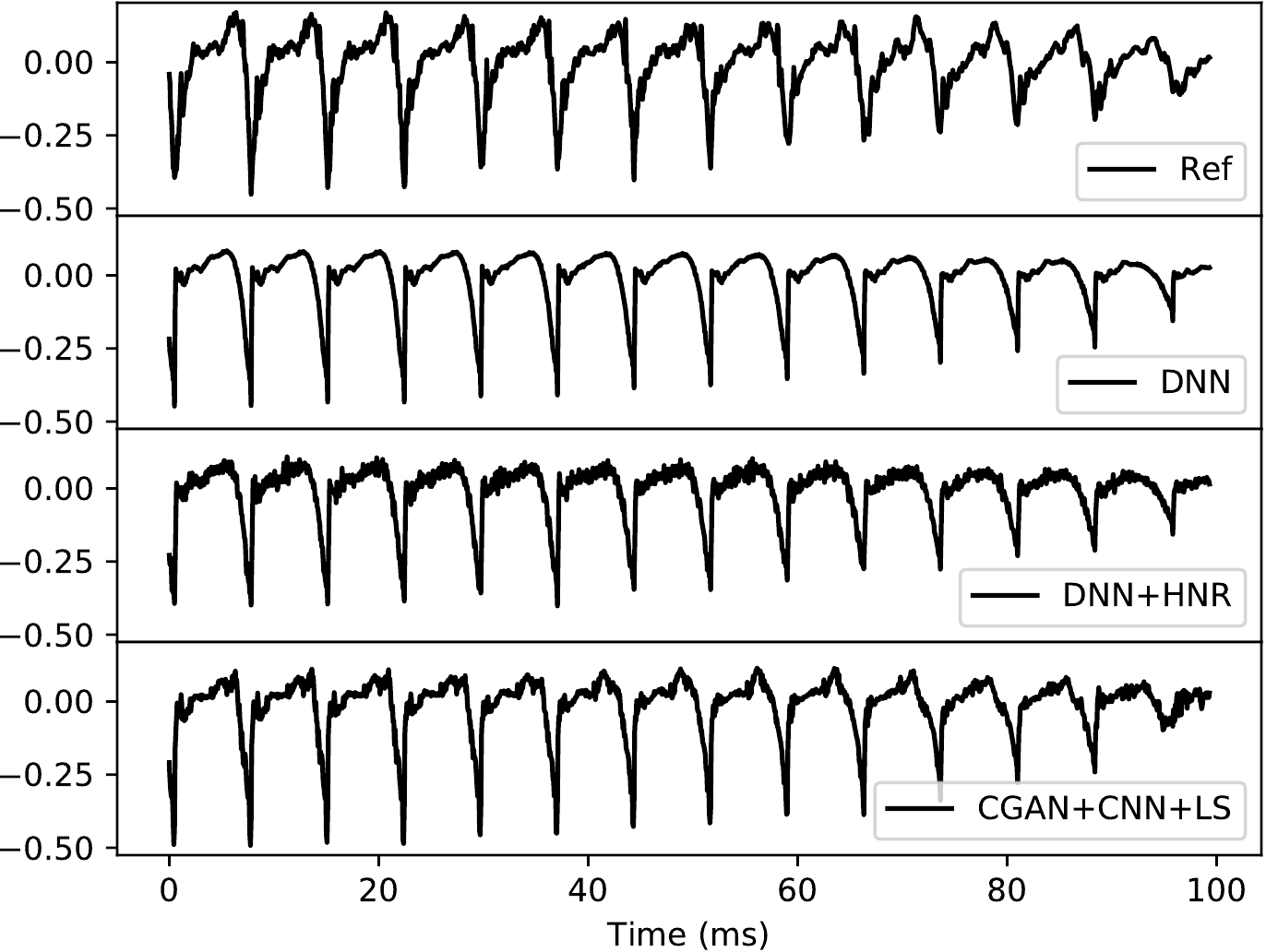}
        \caption{{\it Glottal excitation signals after PSOLA. Ref: Reference excitation signal. DNN: excitation generated using the baseline DNN model. DNN+HNR: baseline DNN with additive shaped noise. CGAN+CNN+LS: convolutive LS-GAN conditioned with acoustic features.
        }}
        \label{fig:excitationwav}
        %\vspace{-0.4cm}
      \end{figure}
      
     \begin{table}[]
          \caption{The objective scores of GANs and DNN. MSE: mean square error, PCC: Pearson correlation coefficient.}
     \label{tab:objscores}
     \centering
        \begin{tabular}{|c|c|c|}
            \hline
             Model                 & MSE &  PCC  \\ \hline
              DNN                   & 0.2458 &  0.86  \\
             GAN             & 0.66   &  0.68  \\ 
             CGAN            & 1.9135 &  0.54  \\ 
             CGAN+CNN            & 0.4469 &  0.76  \\ 
             CGAN+CNN+LS       & 0.4644 &  0.76 \\ \hline
             
        \end{tabular}
 \end{table}
 Figure~\ref{fig:excitationwav} shows the voiced souce excitation signal after pitch-synchronous ovelap-add (PSOLA) \cite{moulines1995-psola}. The excitation signal generated by the baseline DNN is smooth and without a noise component, and therefore
 shaped noise is added to the signal to match the predicted HNR values. 
 %The HNR values were added to incorporate the noise component into the DNN-based excitation signal.
 The GAN-based model, however, is able to generate a noise component similar to the reference waveform without using any HNR-based post-processing.
 
\subsection{Subjective evaluation}
%In glottal vocoding, to increase the quality of synthetic speech the predicted glottal pulses were further processed by acoustic parameters such as line spectral frequencies of voice source spectrum and harmonic-to-noise ratio (HNR) values. 

%In GlottDNN vocoder harmonic-to-noise ratio (HNR) values were added to glottal pulses predicted from DNNs to incorporate the noise component into glottal pulses since the glottal pulses generated by the baseline DNN GL model were conditional averaged across the acoustic features. Figure \ref{fig:excitationwav} shows the glottal excitation signal after post-processing. Top row shows the reference excitation waveform, second row shows the excitation signal created using pulses predicted from the baseline DNN that was very smooth, third row shows the excitation signal after adding the HNR values to the previous excitation signal, and fourth row shows the excitation signal created using pulses generated from GANs. We can observe that GANs able to capture the stochastic component better then the DNNs.

Subjective evaluation was conducted with the comparison category rating (CCR) test \cite{recommendation1996800} between three systems: the baseline DNN (denoted ``DNN''), the baseline DNN with HNR (denoted ``DNN+HNR'') and the GAN-based glottal generation (denoted ``GAN''). Among the GAN-based glottal generation models, we selected the CGAN+CNN+LS system since it performed better than the other systems in informal listening tests. A total of 11 utterances from the test set were randomly selected for the listening test.

A crowd sourcing platform, CrowdFlower \cite{crowdflower}, was employed for the subjective evaluation and followed the same setup as in \cite{Airaksinen2017}. A set of 13 utterances were used as control utterances that included null pairs and anchor samples \cite{recommendation1996800}. Listeners who performed with at least 75 \% accuracy were allowed to participate in the actual listening test. The tests were made available to the English speaking countries, and top four countries in EF English Proficiency Index rankings \cite{efi}. A total of 3850 judgments were made by 50 listeners.

%The results of the listening test are shown in Figure \ref{fig:ranking}. The DNN+HNR method performed better than other two methods. However, the comparison between GAN and the DNN without HNR shows that the former was evaluated to be slightly better. This is a promising result indicating that GANs are able to capture also the stochastic component rather than solely modeling the smooth glottal waveform as done by the DNN.

\begin{figure}[t]
        \centering
        \includegraphics[width=0.8\linewidth, height=4.5cm]{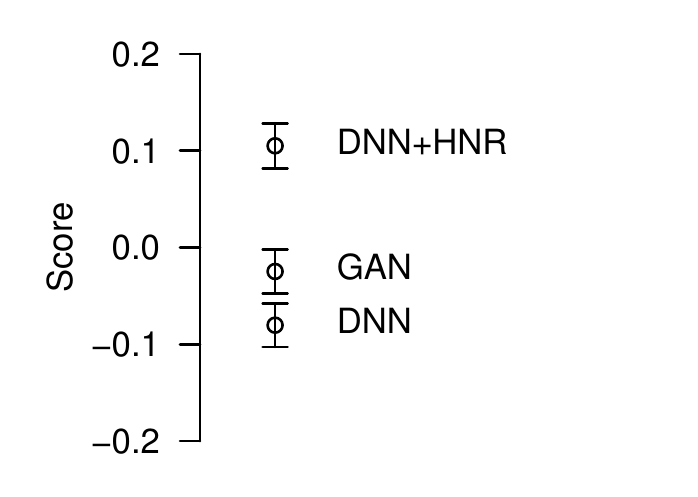}
        % \vspace{-0.4cm}
        \caption{{\it  Subjective listening test results (CCR test) with their 95\% confidence intervals on synthesis quality. }}
        \label{fig:ranking}
        %\vspace{-0.4cm}
      \end{figure}

 The results of the listening test are shown in Figure \ref{fig:ranking}. The DNN+HNR method performed better than other two methods, indicating the perceptual relevance of a stochastic component in excitation. Moreover, in the comparison between GAN and the DNN without HNR, the former was rated slightly higher. This is likely related to GANs ability to generate stochastic variability, rather than producing smooth glottal waveforms as done by the DNN.

  \section{Conclusions} \label{sec:conclusions}
  This study proposed a new method to model glottal excitation waveforms in statistical parametric speech synthesis using generative adversarial networks (GANs). We modified the vanilla GAN in various forms comparing the system performance in generation of glottal pulses. In our experiments, the deep convolutional neural networks -based GANs outperformed the DNN-based GANs. We also compared glottal pulses generated by the GANs with DNNs. The subjective evaluation gave encouraging evidence showing that GANs are able to better reproduce the stochastic component in the glottal excitations than DNNs. The GANs are still relatively new and definitely require more research to understand their full potential in SPSS. 
  
%   \section{Acknowledgements}
  
%     The ISCA Board would like to thank the organizing committees of the past INTERSPEECH conferences for their help and for kindly providing the template files.

  \newpage
  \eightpt
  \bibliographystyle{IEEEtran}
  \bibliography{mybib}
\end{document}